\documentclass{aa}  
\usepackage{graphicx}

\usepackage{txfonts}

\usepackage[usenames,dvipsnames,svgnames,table]{xcolor}

\begin{document}

   \title{The ultra-diffuse galaxy NGC~1052-DF2 with MUSE: \\ II. The population of DF2: stars, clusters and planetary nebulae\thanks{Based on observations collected at the European Southern Observatory under ESO programmes 2101.B-5008(A) and 2101.B-5053(A).}} 
   
   \titlerunning{NGC~1052-DF2: II. The population of DF2}

   \author{J\'er\'emy Fensch\inst{1}, Remco F. J. van der Burg\inst{1}, Tereza Je\v{r}\'abkov\'a\inst{1,2,3}, Eric Emsellem\inst{1,4}, Anita Zanella\inst{1}, Adriano Agnello\inst{1,5}, Michael Hilker\inst{1}, Oliver M\"uller\inst{6}, Marina Rejkuba\inst{1}, Pierre-Alain Duc\inst{6}, Patrick Durrell\inst{7}, Rebecca Habas\inst{8},  Sungsoon Lim\inst{9}, Francine R. Marleau\inst{8}, Eric W. Peng\inst{10,11} and Rubén Sánchez Janssen\inst{12}
          }
          
        \authorrunning{Fensch, van der Burg, Jeřábková, Emsellem, et al.}
	\offprints{J. Fensch (jfensch@eso.org)}

   \institute{
   European Southern Observatory, Karl-Schwarzschild-Str. 2, D-85748 Garching, Germany
   \and
Helmholtz Institut f\"{u}r Strahlen und Kernphysik, Universit\"{a}t Bonn, Nussallee 14–16, 53115 Bonn, Germany
\and
Astronomical Institute, Charles University in Prague, V 
Hole\v{s}ovi\v{c}k\'ach 2, CZ-180 00 Praha 8, Czech Republic
   \and
Universit\'e Lyon 1, ENS de Lyon, CNRS, Centre de Recherche Astrophysique de Lyon UMR5574, F-69230 Saint-Genis-Laval France
   \and
   DARK, Niels Bohr Institute, University of Copenhagen, Lyngbyvej 2, 2100 Copenhagen, Denmark
   \and
   Observatoire Astronomique de Strasbourg  (ObAS), Universite de Strasbourg - CNRS, UMR 7550 Strasbourg, France
   \and
   Youngstown State University, One University Plaza, Youngstown, OH 44555 USA
   \and
    Institut f{\"u}r Astro- und Teilchenphysik, Universit{\"a}t Innsbruck, Technikerstra{\ss}e 25/8, Innsbruck, A-6020, Austria
   \and
   NRC Herzberg Astronomy and Astrophysics Research Centre, 5071 West Saanich Road, Victoria, BC, V9E 2E7, Canada
   \and
   Department of Astronomy, Peking University, Beijing, China 100871
   \and
   Kavli Institute for Astronomy and Astrophysics, Peking University, Beijing, China 100871
   \and 
   UK Astronomy Technology Centre, Royal Observatory, Blackford Hill, Edinburgh, EH9 3HJ, UK
             }

   \date{Received November 12, 2018;}

  \abstract{
  
NGC~1052-DF2, an ultra diffuse galaxy (UDG), has been the subject of intense debate. Its alleged absence of dark matter, and the brightness and number excess of its globular clusters (GCs) at an initially assumed distance of 20~Mpc, suggested a new formation channel for UDGs. We present the first systematic spectroscopic analysis of both the stellar body and the GCs (six of which were previously known, and one newly confirmed member) of this galaxy using MUSE@VLT.  Even though NGC~1052-DF2 does not show any spatially extended emission lines we report the discovery of three planetary nebulae (PNe). We conduct full spectral fitting on the UDG and the stacked spectra of all GCs. 
The UDG's stellar population is old, 8.9$\pm$1.5~Gyr, metal-poor, with [M/H] = $-$1.07$\pm$0.12 with little or no $\alpha$-enrichment. The stacked spectrum of all GCs indicates a similar age of 8.9$\pm$1.8~Gyr, but lower metallicity, with [M/H] = $-$1.63$\pm$0.09, and similarly low $\alpha$-enrichment. There is no evidence for a variation of age and metallicity in the GC population with the available spectra. 
The significantly more metal-rich stellar body with respect to its associated GCs, the age of the population, its metallicity and alpha enrichment, are all in line with other dwarf galaxies. NGC1052-DF2 thus falls on the same empirical mass-metallicity relation as other dwarfs, for the full distance range assumed in the literature. We find that both debated distance estimates (13 and 20~Mpc) are similarly likely, given the three discovered PNe.
}


   \maketitle
%
%
\section{Introduction}

Ultra diffuse galaxies (UDGs) are a particular type of low-surface brightness galaxies, defined as having central surface brightnesses of $\mu_{g,0}>24$~mag.arcsec$^{-2}$, and sizes of R$_\mathrm{eff} > 1.5$~kpc \citep{vanDokkum2015}. Galaxies with such properties were already known for several decades \citep{sandage1984,impey88,dalcanton97,conselice03}, but their particularly high abundance in galaxy clusters drew attention in the last few years \citep[e.g.][]{vanDokkum2015,koda15,mihos15,munoz15,vanderBurg2016}. 
UDGs are now routinely identified also in groups and in the field \citep{roman17,vanderBurg17,Shi2017,Mueller18}. 

To explain their high abundance in over-dense regions, such as the Coma cluster, \citet{vanDokkum2015} proposed that UDGs may be hosted by massive, MW-like, dark matter (DM) halos that could protect them from environmental effects. One UDG in particular, DF44, was measured to have a stellar velocity dispersion consistent with a $10^{12}$~M$_\sun$ halo \citep{vanDokkum2016}. In addition, the empirical linear relation observed between the mass of the globular cluster (GC) system and the halo mass \citep{Blakeslee1997, peng2004, harris17} allows one to use this quantity to assess the DM content of UDGs. The high number of GCs around DF44 ($\sim$ 100) would confirm the hypothesis of it being hosted by a very massive DM halo, along with few other UDGs with a GC excess. But most UDGs have GC systems typical of dwarf galaxy DM halos \citep{beasley2016tr, amorisco18, lim18}. This is in line with a stacked weak-lensing study performed by \citet{Sifon2018}, showing that not all UDGs can have halo masses similar to those estimated for DF44. Formation scenarios need to explain how galaxies with similar masses and morphologies may be hosted in a broad variety of DM halo masses. 

\citet{dicintio17} suggested the possibility that internal processes (i.e. gas outflows associated with feedback) can, under some circumstances, kinematically heat the distribution of stars and form very extended systems similar to UDGs. 
An early cessation of star formation at $z \sim$ 2 would render their stellar masses, and associated surface brightness, low \citep[cf.][]{Yozin2015}. At different quenching times, such a scenario results in UDG-like galaxies with low metallicities ($-$1.8$\lesssim$ [Fe/H] $\lesssim$ $-$1.0) and a range of ages \citep{Chan2018}. This is supported by photometric \citep{Pandya2018} and spectroscopic observations \citep{ Kadowaki2017,Gu2018,Ferre-Mateu2018, Ruiz-Lara2018} of UDGs. The general consensus is that UDGs have stellar populations that are typically old (> 9~Gyr) and metal-poor ($ \big[ \mathrm{M} / \mathrm{H} \big] \sim $ $-$0.5 to $-$1.5). Moreover, these studies found that the UDGs' stellar masses and stellar metallicities fall on the empirical relation found for dwarf galaxies \citep{Kirby2013}. They conclude that UDGs are most likely the result of both internal processes, such as bursty star formation histories (SFH) or high-spin halos \citep{Amorisco2016, Rong2017}, and environmental effects such as tidal disruption \citep{Collins2013, Yozin2015}. One may note that UDGs may also form in tidal debris \citep[see e.g.][]{Kroupa2012,Duc2014, Bennet2018}.

To reconcile both the discovery of UDGs with exceptional characteristics such as DF44 and the average properties of typical UDGs, several different formation channels need to be invoked. Most stellar population studies have targeted ``ordinary'' UDGs, with typical dwarf galaxy DM haloes. While such galaxies may be well represented in current hydrodynamical simulations \citep[see e.g.][]{Chan2018}, an open question is how more extreme cases (for instance UDGs with an extremely high, or low, halo mass for their stellar mass) have formed. 

Of particular recent interest is the UDG NGC~1052-DF2\footnote{This name was tagged by \citet{vanDokkum2018_nat}. However, we note that this galaxy was already identified under different names (KKSG04, PGC3097693, [KKS2000]04, see e.g. \citealt{Karachentsev2000, Trujillo2018}). We choose to keep NGC~1052-DF2 as the galaxy has been popularized under this name in the recent literature.} (hereafter DF2), which may have a special formation channel. Using the velocities of 10 GCs associated with DF2, \citet{vanDokkum2018_nat} claimed a low total mass that is consistent with the stellar mass only \citep[however see][for a re-analysis]{Martin2018, Famaey2018, Laporte2018}. Hypotheses put forward by \citet{vanDokkum2018_nat}  suggest that DF2 may have been formed by gas ejected by tides following a merger or quasar winds from the massive elliptical NGC~1052, whose projected distance is only 14$\arcmin$, or $\sim$80~kpc at 20~Mpc distance.

A second striking feature of this galaxy is its GC system. DF2 has 12 confirmed GCs (Emsellem et al., submitted), an unusually large population when compared to normal dwarf galaxies \citep{lim18, amorisco18}. This is at odds with the DM deficiency, as explained above. These GCs are also very luminous: their absolute magnitudes are similar to those of the most massive Milky Way GCs at an assumed distance of 20~Mpc \citep{vanDokkum2018_nat}. \citet{Trujillo2018} advocated for a closer distance of 13~Mpc, and showed that DF2 and its GCs would then fall on the same empirical relation as other UDGs. The exact distance of DF2 is still debated \citep{vanDokkum2018_dist}. The new GC candidates associated with DF2 from \citet{Trujillo2018}, if confirmed, would move the peak of the GC luminosity function towards fainter magnitudes and alleviate the issue of `too bright' GCs. This would further increase the discrepancy between the DM halo mass estimated through GC kinematics and that from the GC abundance. In a companion paper (Emsellem et al., submitted, hereafter Paper~\textsc{I}), we have indeed confirmed one new candidate GC from \citet{Trujillo2018}.

In this series of two papers, we study DF2 with MUSE observations taken at the VLT. Thanks to the field of view of this integral field spectrograph, we are able to simultaneously probe the stellar body of the UDG and seven bright associated GCs, for the first time. While Paper~\textsc{I} focuses on the kinematics of the UDG, this paper presents a stellar population analysis of this galaxy and its associated GCs. 

In Section~\ref{data} we present the data reduction, sky removal and extraction of spectra. We estimate the age and metallicity of the stellar body and the GCs in Section~\ref{sec:estimationparameters}. We report the discovery of three planetary nebulae in Section~\ref{sec::PN}. We discuss the origin and the distance of the UDG and its association with the surrounding GCs in Section~\ref{discussion}. The conclusions are in Section~\ref{conclusion}.

\section{Data}
\label{data}
The details of the observations, reduction and flux extraction procedures are detailed in Paper~\textsc{I}. In the following we summarize the main points of the procedure.

\subsection{Observation \& Reduction}

\begin{figure}[ht]
   \centering
   \includegraphics[angle=0,width=9cm]{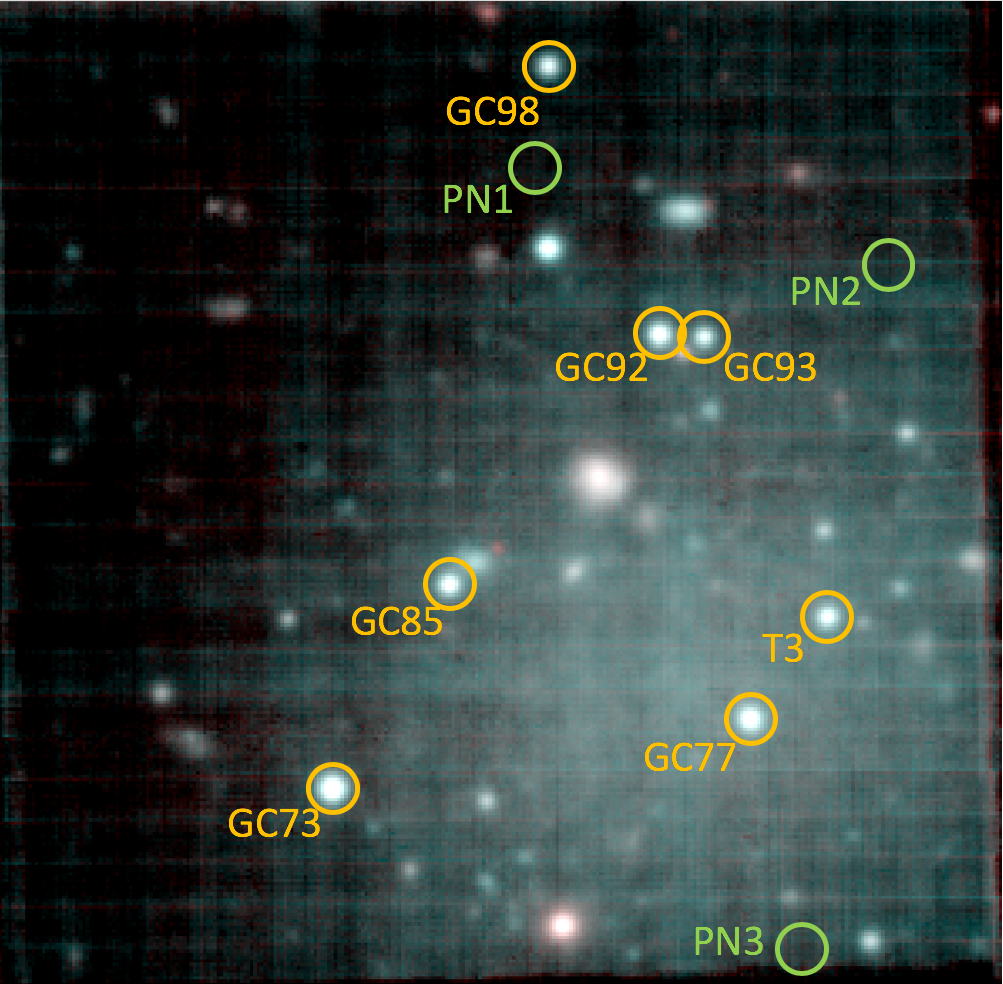}
      \caption{Mock true color image of the collapsed MUSE cube within the F606W and F814W HST bandpasses. The GCs confirmed in Paper I are shown with orange circles. The location of the  three confirmed PNe is shown with green circles. The PNe do not show up in these mock broadband images. The field of view is 1$\arcmin\times 1\arcmin$.
              }
         \label{fig:UDG}
   \end{figure}

MUSE observations of NGC1052-DF2 were conducted via two ESO-DDT programs (2101.B-5008(A) and 2101.B-5053(A), PI: Emsellem) between July and November 2018 amounting to a total of $\sim$5.1h on-target integration time. We obtained 28 individual exposures with slight dithers and rotations to account for systematics due to the slicers. We deliberately offset the MUSE field by $\sim 8''$ with respect to the centre of the galaxy (cf.~Fig.~\ref{fig:UDG}) to include an area where the surface brightness of the UDG is several magnitudes fainter than in the centre, which is used for the sky removal.

The OBs were all reduced using the latest MUSE \textsc{esorex} pipeline recipes (2.4.2). The reduction follows the standard steps. As the object is very faint and standard sky subtraction was not sufficient to recover a good quality signal, the full sky subtraction was done with the principle component analysis-based software Zurich Atmosphere Purge (\texttt{ZAP}, \citealt{Soto2016}). The principal components, or eigenspectra, are derived from the outermost regions of the MUSE object cube, where the sky is most dominant: the sky region is defined by excluding the bright sources and an ellipse centered on the UDG (see Paper I). In the following, we use as fiducial datacube the output of the \texttt{ZAP} procedure with an ellipse of circularized radius 30$\arcsec$, 45 eigenvalues and 50 spectral bins for the continuum filter. We discuss the effect of these parameters on the results in Section~\ref{sub:params}. The final data set, rendered in a mock HST broadband color image, using the same filters as in \citet{vanDokkum2018_nat}, is shown in Fig.~\ref{fig:UDG}.

\subsection{Extraction of spectra}

The detection of the sources is described in Paper I.
 
We first create a spatial mask, presented in Paper I, to remove the background and foreground objects surrounding the UDG. We extract the spectrum of the UDG by summing each channel of the masked cube with a spatial weight corresponding to the flux of the UDG in the \emph{HST} F814W image. 

The GC and PN spectra are extracted with a Gaussian weight function to provide a S/N-optimized extraction. The full width at half-maximum is set to $\sim 0.8$'' to approximately match the point spread function. The background is measured locally with identical apertures in eight nearby locations that do not overlap with identified sources. In each channel we obtain the source flux by subtracting the median of the sky exposures from the weighted sum of the source spectrum. The dominant source of uncertainty is taken from the scatter in the sky spectrum values. The relative velocities are small (see Paper I), thus we do not correct for the relative velocities of the GCs. Contrarily to \citet{vanDokkum2018_gc}, we do not weight each GC by their SNR. This would provide us with the highest reachable SNR, but the brightest source, GC73 (see Fig.\ref{fig:UDG}), would dominate the stack.

To estimate the physical spread in the different parameters (age, [Fe/H], $\alpha$-enrichment) in the GC population we also create 100 bootstrapped spectra. These are new spectra constructed by adding together seven GC spectra that are randomly picked from the sample with replacement. 

\section{DF2's stellar populations: stellar body and GCs}
\label{sec:estimationparameters} 
   \begin{figure*}[h]
   \centering
   \includegraphics[angle=0,width=17cm]{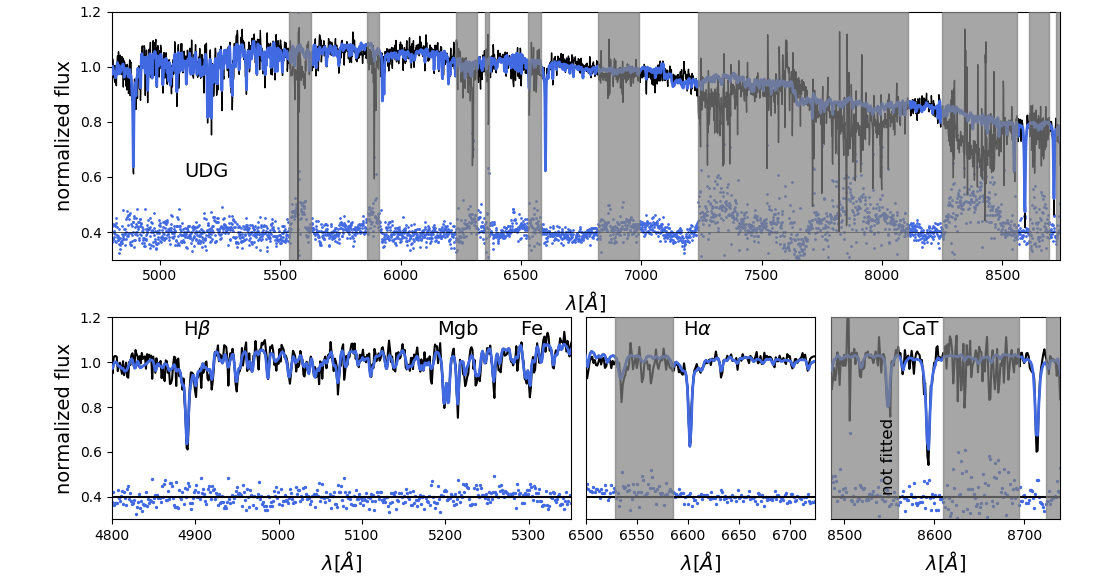}\\
   \includegraphics[angle=0,width=17cm]{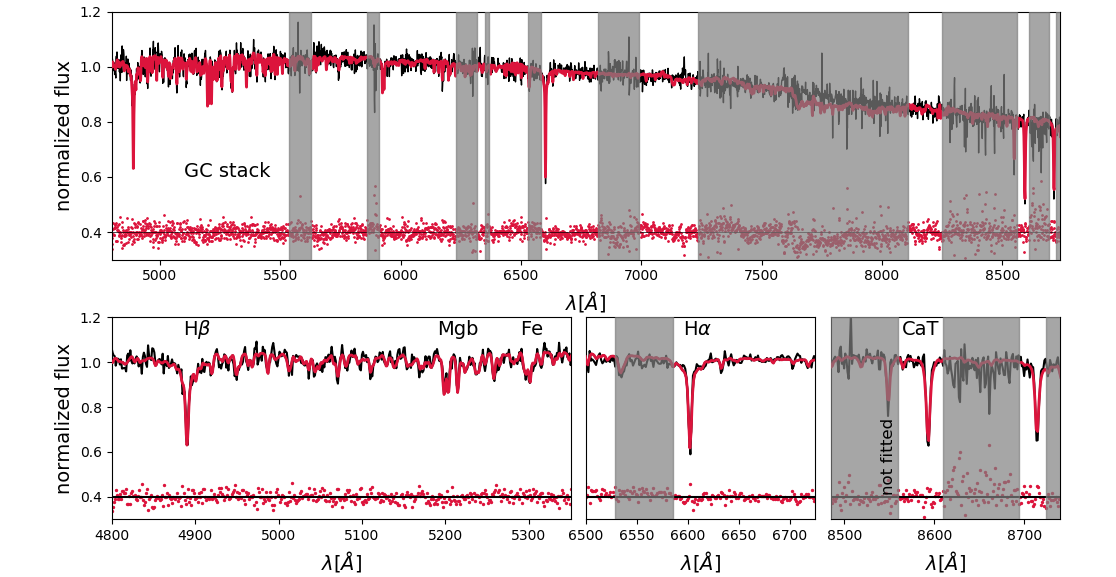}
      \caption{Comparison between the spectrum and the best fit from pPXF for the UDG and the GC stack spectra. The three plots in the bottom part of each panel show zooms on the important absorption lines. The gray regions are not taken into account for the fit. The scatter points show the residuals. 
              }
         \label{fig:best_fit}
   \end{figure*}

We show in Fig.~\ref{fig:best_fit} the spectrum obtained for the UDG and the stack of all GCs. We note strong Balmer and calcium triplet (CaT) absorption lines, plus shallower absorption lines such as Mg and Fe. We do not detect any emission lines. This is consistent with the non-detection of atomic gas which implies a stringent upper-limit on the gas fraction of DF2 \citep[below 2\%, see ][]{Chowdhury2018}. Around the H$\alpha$ line, we estimate a signal-to-noise ratio (SNR) of 62~pix$^{-1}$ for DF2 and 72~pix$^{-1}$ for the stack of all GCs (see Paper I). 

\subsection{Fitting procedure}
\label{sub::fitting}
We use the fitting routine pPXF \citep{Cappellari2004, Cappellari2017} combined with the eMILES library \citep{Vazdekis2016}. The details of the fitting procedure are given in Paper I. In the following we summarize the main points of the procedure.

As template spectra we use the eMILES single stellar populations (SSPs) with a \citet{Kroupa2001} initial mass function (IMF) and the Padova 2000 \citep{Girardi2000} isochrones which were shown to perform well in the expected regime of old and low metallicity stellar populations \citep{Conroy2009}. The original range of metallicity values being rather sparse (only seven metallicity covering $\big[ \mathrm{Fe} / \mathrm{H}\big]$ from $-$2.32 to 0.22 with logarithmic spacing), we linearly interpolate for sixteen more metallicity values, between $\big[\mathrm{Fe} / \mathrm{H}\big]=-$2.32 and $-$0.71, following \citet{Kuntschner2010}. 

To avoid being biased by the flux calibration differences between our MUSE data and the eMILES library, we make use of multiplicative polynomials during the fit \citep{Cappellari2017}. For the study we chose to allow for a 12-degree Legendre multiplicative polynomial and the impact of changing the degree is discussed in Sect.~\ref{sub:params}.

\subsection{Ages and metallicities - Fitting method}
\label{sub:params}

   \begin{figure}
   \centering
   \includegraphics[angle=0,width=10cm]{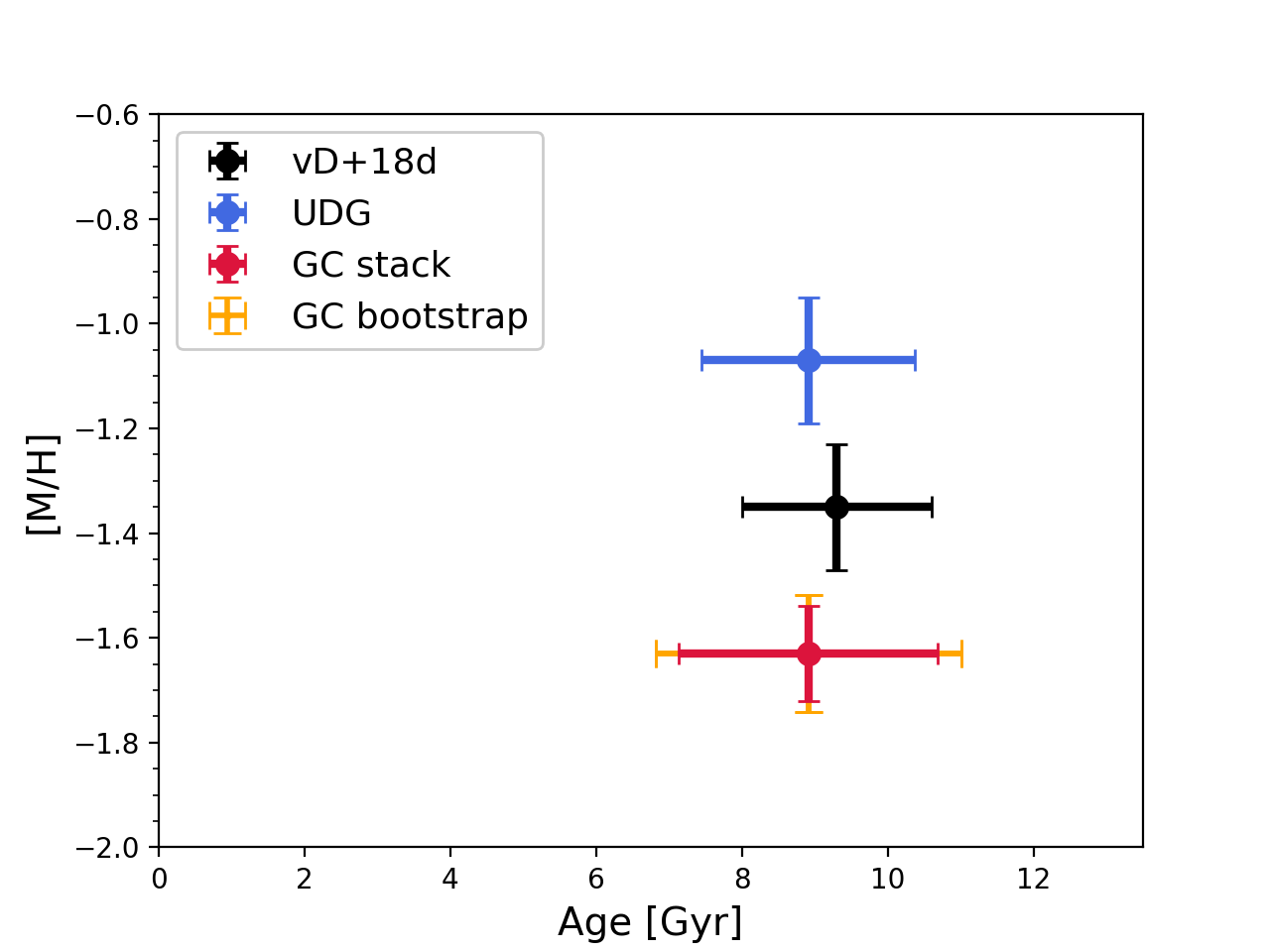}
      \caption{Location of the best fit and 1-$\sigma$ error bars in the age-metallicity plane for the UDG and the GC stack. The result of the study by \citet{vanDokkum2018_gc} is shown in black. In orange is shown the location of the median age and metallicity of the GC bootstrap sample. Note that the orange error bar is both a measure of the error of the fit and an estimate of the physical parameter spread intrinsic to the GC sample.
              }
         \label{fig:age_Z}
   \end{figure}

We estimate the stellar population parameters by fitting single stellar populations (SSPs) to our spectra. The parameter uncertainties are derived from fitting 100 new spectra, constructed by adding the randomly shuffled residuals to the best fit. The best SSP fits for the UDG and GC stack are shown in Fig.~\ref{fig:best_fit}. It should be noted that the first CaT line is masked during the fit \citep[as in][]{vanDokkum2018_nat}, because it is located in a region affected by sky residuals. It is nonetheless well recovered by the fits, which independently shows that the sky subtraction did not affect these lines.

The location of the UDG and the GCs in the age-metallicity plane is shown in Fig.~\ref{fig:age_Z} with the estimation of age and metallicity for the stack of GCs from \citet{vanDokkum2018_gc}, along with their 1$\sigma$ error bars. For the UDG we find a best fitting age of $8.9 \pm 1.5$~Gyr and metallicity [M/H]=$-$1.07$\pm$0.12. For the full stack of GCs, we find a best fitting age of $8.9 \pm 1.4$~Gyr and metallicity of  [M/H]$= -1.63 \pm 0.09$. The parameters of the best fits are not sensitive to a change of the degree of the multiplicative polynomial between 11 and 15, nor to a change of parameters in the \textsc{ZAP} procedure (masked radius of 30$\arcsec$ or 36$\arcsec$, number of eigenvalues of 30, 45 or 50, and continuum filter window size of 30 and 50~$\AA$). 

It should be noted that our method does not consider the detailed continuum shape to derive the parameters, because of the use of multiplicative polynomials. In order to check the consistency of our estimates with the broad-band colors, we compute the AB magnitude color of the eMILES templates in F606W$-$F814W. The color of both best fit templates, respectively 0.40~mag for the UDG's and 0.35~mag for the GC stack's, agree with the colors computed by \citet{vanDokkum2018_gc}: respectively 0.37$\pm$0.05~mag and 0.35$\pm$0.02~mag for the UDG and the GC stack. 

The age and metallicity estimated for the GC stack are consistent within 1-$\sigma$ for the ages and 2-$\sigma$ for the metallicity to the values obtained by \citet{vanDokkum2018_gc}: age of $9.3^{+1.3}_{-1.2}$~Gyr and [Fe/H]~$= - 1.35 \pm 0.12$. We obtain a lower metallicity for our GC stack, but it should be noted that we do not use the same stellar libraries for the fits, and the IMF they assume is not described. Furthermore, the spectral region studied in \citet{vanDokkum2018_gc} extends further into the blue compared to our MUSE data, where different spectral diagnostics contribute to the fit. Even though these differences may drive systematic shifts between the parameters measured in different studies, we note that in this work we make a direct comparison between the GCs and the stellar body from a single data set, which have similar SNR and assumptions. Even though the exact age and metallicity may be affected by different systematics, the relative differences between GCs and stellar body are significant and robust.

To quantify the spread in age and metallicity inside the GC population, we use our method on the bootstraped spectra. The median of the 100 realisations has the same age and metallicity as the GC stack. The error bar shows the propagation of the measurement error and the dispersion of the results in the bootstrap sample. We obtain an age of $8.9 \pm 2.1$ and a metallicity of $-1.63 \pm 0.11$. We see that it is of the same order of magnitude as the error on the estimation of the parameters of the bootstrap, meaning that there is no significant evidence for a spread in properties between the individual clusters. Finally, we used pPXF on three radial sectors of the UDG: inside 0.5 $R_{\mathrm{e}}$, between 0.5 and 1 $R_{\mathrm{e}}$ and between 1 and 1.5 $R_{\mathrm{e}}$, where $R_{\mathrm{e}}$ is the effective radius of DF2 (see Paper I). The best fits have all the same age, but the central sector's metallicity estimate  has a higher metallicity: [Fe/H] = -1.07$\pm$0.12 compared respectively to [Fe/H]= -1.19$\pm$0.12 and -1.19$\pm$0.14 for the two outer sectors. We thus note that the higher metallicity found in the center of DF2 is not significant and that the metallicity gradient in DF2 is consistent with being flat.

\subsection{Ages and metallicities - Spectral indices}

   \begin{figure*}
   \centering
   \includegraphics[width=16cm]{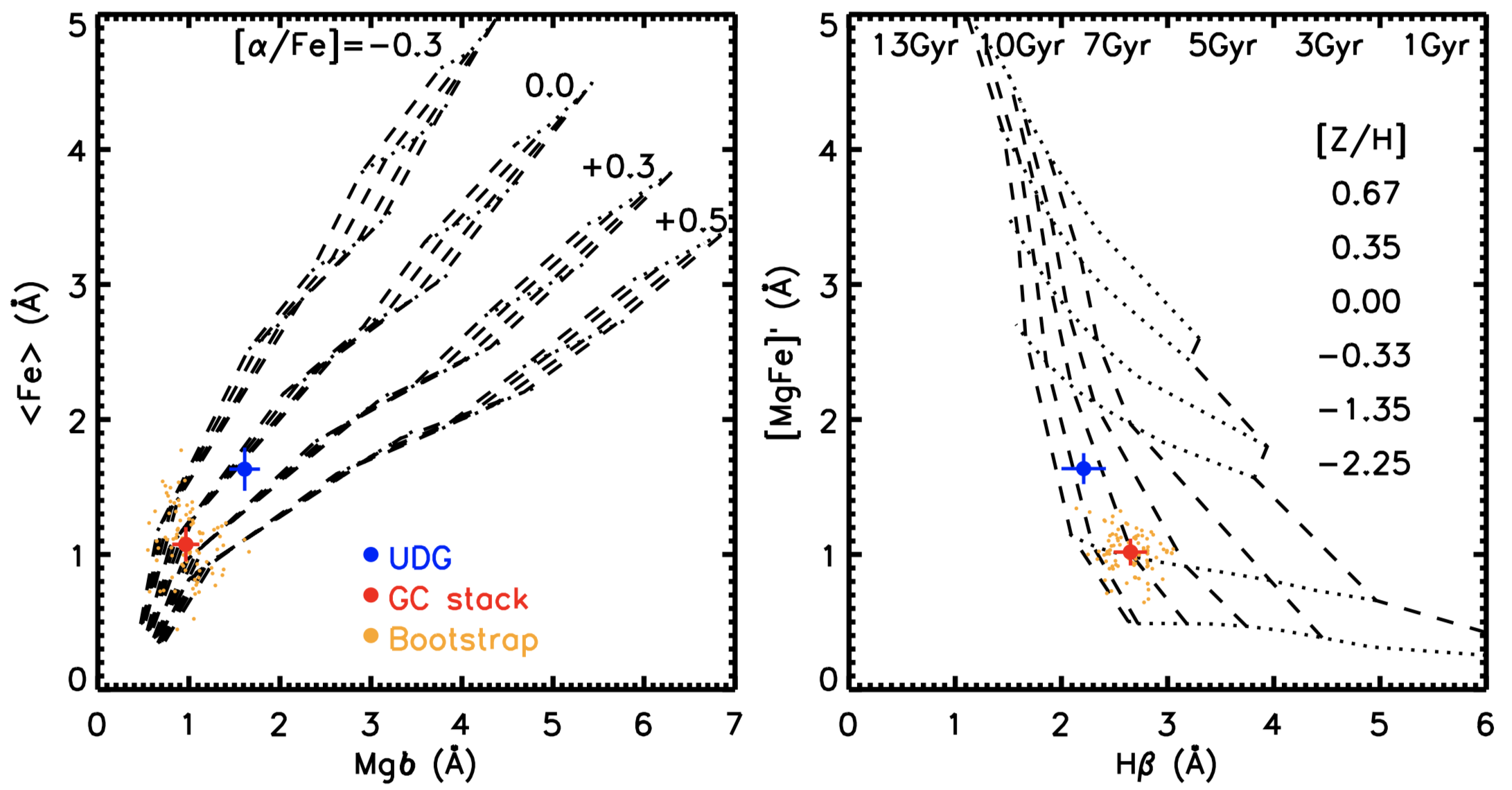}
   
      \caption{Lick/IDS indices measured from our spectra of the UDG (blue) and the stacked GC spectrum (red). The orange cloud of 100 points show the bootstrap runs of the GCs, as detailed in the text, and indicate the scatter intrinsic to the stack. The underlying model grid is based on \citet{thomas10}. A grid with ages of {1, 3,} 5, 7, 10 and 13~Gyr (dashed lines), and [Z/H] of $-2.25$, $-1.35$, $-0.33$, 0.0, $+0.35$, and $+0.67$~dex (dotted lines) is over-plotted on both panels. For clarity, lines for 1 and 3~Gyr are omitted in the left panel. \textit{Left:} comparison of Mg$\,b$ and $<{\rm Fe}>$ as a measure of $\alpha$ enrichment. \textit{Right:} comparison of $[\mathrm{MgFe}]'$, a measure of total metallicity, and H$\beta$, which is primarily an age indicator.   }
              
         \label{fig:lick_comparison}
   \end{figure*}

\begin{table*}

\caption{Lick/IDS indices for the UDG- and stacked GC spectra in $\AA$. First errors represent statistical uncertainties measured from the (stacked) spectra. For the GCs we quote, in addition, the scatter obtained from bootstrapping the GCs that end up in the stack.}   
\label{table:lick}     
\centering          
\begin{tabular}{l c c c c c c}     
 \hline \hline 
&H$\beta$& Mg$b$ & Fe 5270& Fe 5335& <Fe>& $[\mathrm{MgFe}]'$\\
\hline 
UDG & $2.21\pm0.21$&$1.61\pm0.17$ & $1.69\pm0.20$ & $1.57\pm0.25$ & $1.63\pm0.16$ & $1.64\pm0.11$ \\
GCs & $2.65\pm0.16\pm0.19$&$0.97\pm0.15\pm0.22$ & $1.07\pm0.18\pm0.18$ & $1.08\pm0.18\pm0.30$ & $1.08\pm0.13\pm0.23$ & $1.02\pm0.10\pm0.14$ \\
\hline 
\end{tabular}
\end{table*}

We use a complementary method to estimate ages and metallicities based on the measurement of spectral line indices. In the following we work in the standardized Lick/IDS system \citep{worthey94}, and we list several key diagnostics in terms of age and metallicity in Table~\ref{table:lick}. Two diagnostics are shown in Fig.~\ref{fig:lick_comparison}, with over-plotted grids of theoretical Lick indices of SSPs, based on the MILES spectral library, are obtained from \citet{thomas10}. 

To study the $\alpha$-enrichment of the GCs and the UDG, we plot in the left panel of Fig.~\ref{fig:lick_comparison} Mg$\,b$, as a probe of the $\alpha$ elements, and $\langle$Fe$\rangle$ \citep[the average of Fe $\lambda$5270 and Fe $\lambda$5335, following][]{evstigneeva07}. The $\alpha$-enrichment of the GC stack and the UDG are not well constrained due to the small separation of iso-[$\alpha$/Fe] lines in the metal-poor regime. 
Still, the diagnostics infer slight $\alpha$-element enrichment: from 0 to 0.15 for the UDG and from 0 to 0.3 for the stack of GCs. This latter value is consistent with the value derived by \citet{vanDokkum2018_gc} ([$\alpha$/Fe]~$ = 0.16 \pm 0.17$). 
 
The right panel of Fig.~\ref{fig:lick_comparison} shows the age-sensitive index H$\beta$ versus $[\mathrm{MgFe}]'=[\mathrm{Mg} b\times \mathrm{(0.72~Fe\lambda 5270 + 0.28~Fe \lambda 5335)}]^{1/2}$ \citep[which probes the total metallicity, following][]{evstigneeva07}. The Lick index suggests a slightly higher metallicity than our full spectral fitting method indicates. However, it confirms the trend given by the first method that DF2 and the stack of all GCs have similar ages but that the UDG has a higher metallicity. 

Even though measurements of Lick indices from individual GC spectra are imprecise given the noise in our spectra, we indicate the cluster-to-cluster variation by showing a distribution of bootstrapped stacks. The scatter of these realizations is represented by the second error quoted in Table~\ref{table:lick}. We note that the bootstrap uncertainties are similar to the formal statistical uncertainties (first errors quoted). Since the bootstrap error, by construction, also includes the statistical uncertainty on the measurement, them being of similar magnitude confirms that there is no significant spread in the properties of the individual clusters.\\

The two different methods indicate that DF2's stellar population has the same old age as the GC's, around 9~Gyr, and is significantly more metal-rich, by around 0.5~dex. 

\section{DF2's planetary nebulae}
\label{sec::PN}

   \begin{figure}
   \centering
   \includegraphics[width=9.5cm]{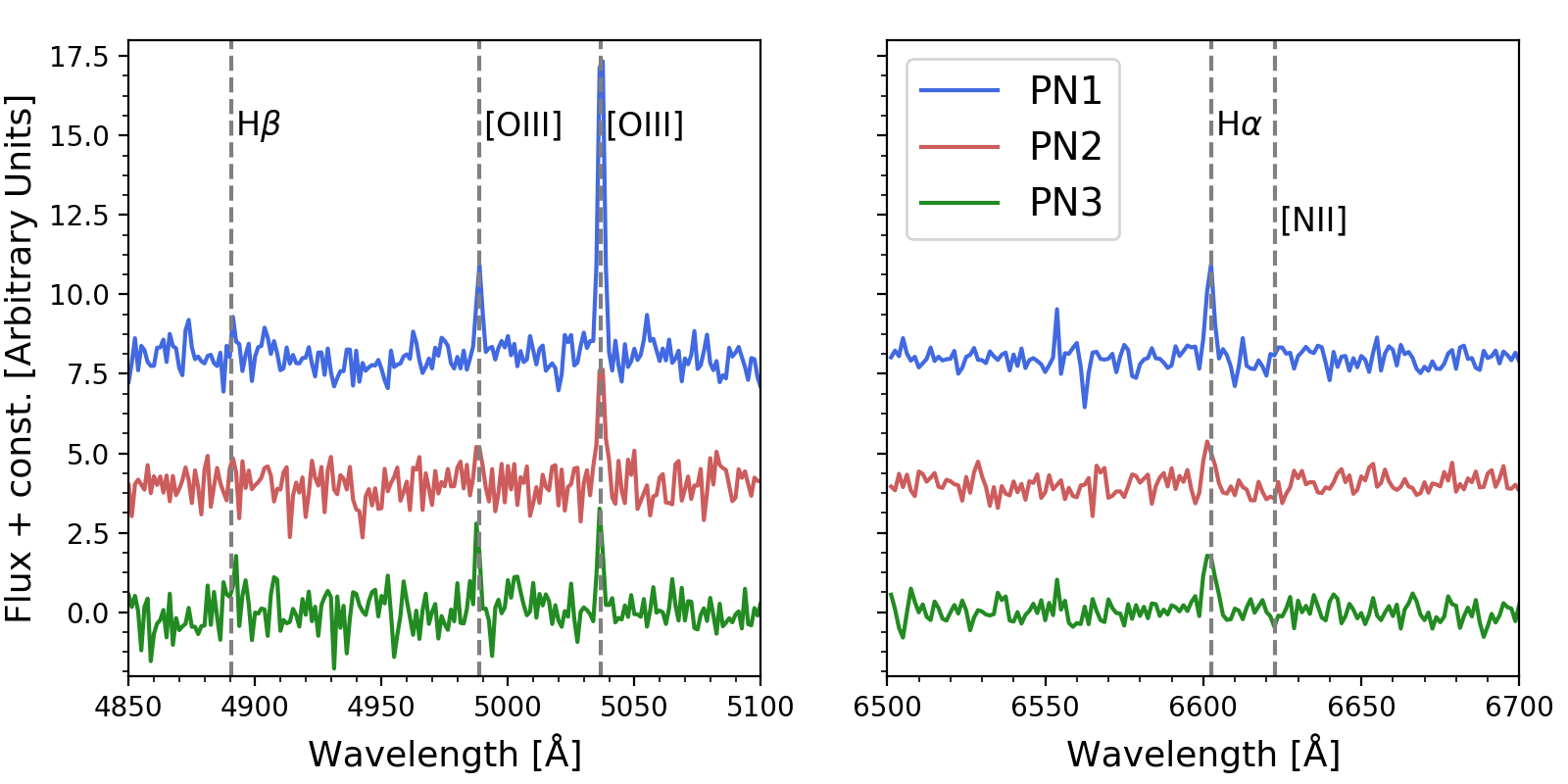}
      \caption{Zoom on the spectra of the three PNe. The left panel shows the region around the [{\sc O~$\,$iii}] doublet and the right panel shows the region around the $\mathrm{H}\alpha$ line. The fluxes are arbitrarily shifted for clearer visualization. The location of the brightest expected lines for PNe, observed for z = 0.06, are shown with labeled vertical dashed lines. }
         \label{fig:PNe}
   \end{figure}

The spectra of the three detected PNe are shown in Fig.~\ref{fig:PNe}. Their kinematic association with the stellar body of the UDG is confirmed in Paper I. We see strong emission from the [{\sc O~$\,$iii}] doublet and $\mathrm{H}\alpha$ lines. However, $\mathrm{H}\beta$ and [{\sc N~$\,$ii}] are not detected for any of the PNe, which prevents us from computing their intrinsic extinction or metallicity. The measurement of their apparent 5007$\AA$ magnitude is given in Table~\ref{table:PNe}. It is defined\footnote{The standard magnitude zeropoint for PNe is set at 13.74 to approximate the absolute V-band magnitude one would observe if all the [{\sc O~$\,$iii}] line emission were distributed over the V-band \citep{Allen1973}.} as:

\begin{equation}
    m_{5007} = -2.5 \mathrm{log}F_{5007} - 13.74
\end{equation}

with $F_{5007}$ is the integrated flux in the second [O~\textsc{iii}] line in erg~s$^{-1}$~cm$^{-2}$. We check our flux calibration by comparing the flux of our GCs with those presented in \citet{Trujillo2018}, with their HST observations. We found that the flux from the MUSE cubes are brighter by $0.064 \pm 0.079$~mag. We neglect this calibration difference and use the flux calibration from MUSE.

We assume a foreground extinction\footnote{Model accessible at https://irsa.ipac.caltech.edu/applications/DUST/} of $0.076 \pm 0.006$~magnitude, corresponding to mean of the computed extinction for the line-of-sight of NGC~1052 \citep{Schlegel1998, Schlafly2011}. The uncertainty includes a propagation of the uncertainty on the foreground extinction, the flux calibration and on the flux measurement. The latter is obtained by re-noising the spectrum with a Gaussian noise with a dispersion measured in the continuum red-ward of the [O~\textsc{iii}] line. H$\beta$ is not detected in any of the PNe and the SNR prevents us to infer a meaningful lower limit on the extinction. Thus, we did not correct for internal extinction. 

\begin{table}
\caption{5007$\AA$ apparent and absolute magnitude for the three PNe. The absolute magnitude is obtained for two assumed distances to DF2: 13 and 20~Mpc.}   
\label{table:PNe}     
\centering          
\begin{tabular}{c c c c}     
\hline \hline
   & $m_{5007}$ & $M_{5007}$  & $M_{5007}$           \\
   &            & at 13~Mpc   & at 20~Mpc            \\  \hline 
PN1  & 28.4 $\pm$ 0.05 & -2.24 $\pm$ 0.05 & -3.18 $\pm$ 0.05  \\
PN2  & 29.32 $\pm$ 0.14 & -1.32 $\pm$ 0.14 & -2.26 $\pm$ 0.14 \\ 
PN3 & 29.91 $\pm$ 0.16 & -0.73 $\pm$ 0.16 & -1.67 $\pm$ 0.16  \\
\hline
\end{tabular}

\end{table}

\section{Discussion}
\label{discussion}

\citet{vanDokkum2018_nat} argued that DF2 was DM-deficient and a very different system from other galaxies, in particular from UDGs that were routinely shown to be hosted by dwarf to MW sized DM halos (see Introduction). In the following we discuss whether DF2 also stands out in terms of its stellar populations.

\subsection{How does DF2 compare with other UDGs?}
    \subsubsection{In terms of stellar populations}
    \label{sub:disc_stel_pop}

   \begin{figure}
   \centering
   \includegraphics[angle=0,width=10cm]{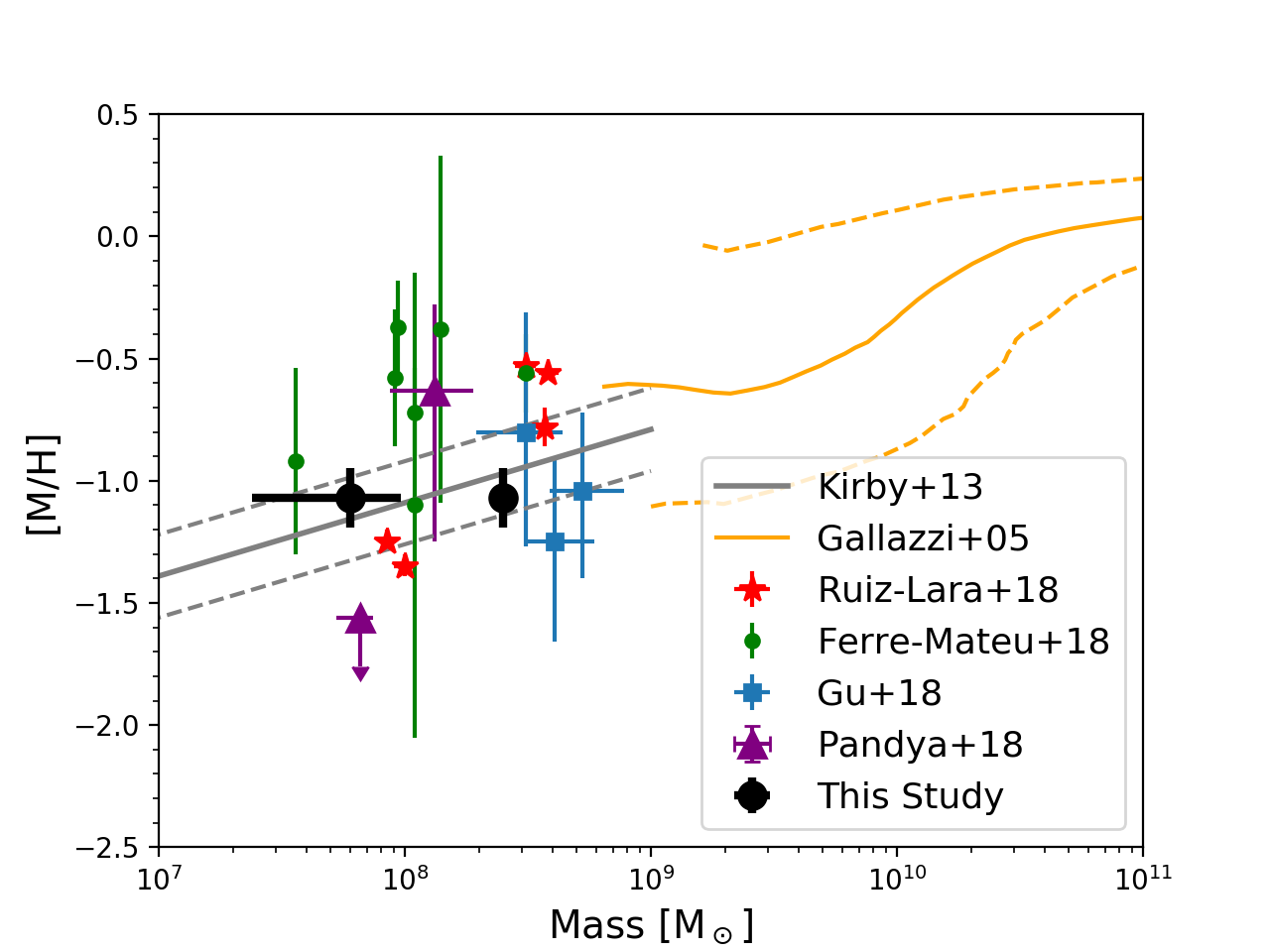}
      \caption{Stellar mass and metallicity of DF2 and UDGs from the literature \citep{Gu2018,Ferre-Mateu2018, Ruiz-Lara2018, Pandya2018}. We show the location of DF2 for two mass estimates, corresponding to the two distance estimates of 13 and 20~Mpc (see text). The empirical mass-metallicity relations for low-mass and high-mass systems (from \citealt{Kirby2013} and \citealt{Gallazzi2005}, respectively) are shown in gray and orange.
              }
         \label{fig:age_Z_litt}
   \end{figure}

In Section~\ref{sub:params}, we estimated the age, the metallicity and the $\alpha$ element enrichment of DF2 and its GCs.

We found that the stellar population of DF2 is old, around 9~Gyr. It should be noted that our age estimate should be taken as a lower-limit, as blue horizontal branch stars could bias our age estimate to lower ages \citep{Schiavon2007, Conroy2018}. This age estimate is however similar to those obtained for other quiescent UDGs \citep{Gu2018, Ruiz-Lara2018, Ferre-Mateu2018}. 

The few UDGs with $\alpha$-enrichment measurements, which are located in the Coma cluster, have [$\alpha$/Fe] estimated to be between 0 and 0.6~dex \citep{Ruiz-Lara2018, Ferre-Mateu2018}. Our estimation of [$\alpha$/Fe] for DF2, between 0 and 0.3 is located within this range of values.

To study the metallicity of DF2, we show in Fig.~\ref{fig:age_Z_litt} the location of the UDG in the mass-metallicity plane along with data from previous studies of quiescent UDGs. We indicate two different stellar masses for DF2: the one inferred for a distance of 20~Mpc \citep[$2-3\times 10^8 \mathrm{M}_{\odot}$; ][]{vanDokkum2018_nat} and the other for a distance of 13~Mpc \citep[$6 \pm 3 \times10^7 \mathrm{M}_\odot $; ][]{Trujillo2018}. We see that DF2 has a similar metallicity as the other UDGs previously studied, and falls on the empirical relation for dwarf galaxies from \citet{Kirby2013} for both mass estimates. We note that our data provide us with a much tighter metallicity estimate than most of the ones available in the literature.

This stellar mass-metallicity relation is interpreted as an effect of self-enrichment. The more massive a galaxy, the less metals are lost to galactic winds launched by star-formation feedback \citep{Kirby2013}. The mass-metallicity relation may also result from the galaxy-wide stellar IMF becoming systematically top-lighter with decreasing baryonic mass or star formation rate as shown to be the case using the IGIMF theory \citep{Koeppen2007, Recchi2015}.

If one assumes that DF2 is DM-deficient \citep{vanDokkum2018_nat}, one would then expect DF2 to be an outlier of the relation, with a lower metallicity than galaxies with the same stellar mass, which typically have a halo mass of $10^{10}$~M$_\odot$ \citep[see e.g.][]{Read2017}. However we see that, for the assumed distance of 20~Mpc, corresponding to the DM-deficiency hypothesis, DF2 lies within the scatter of the relation. Even though the scatter of the relation is quite large ($\sim$ 1~dex), DF2 has a higher metallicity than DF44 ([Fe/H] = $-$1.3 $\pm$ 0.4, see \citealt{Gu2018}) which has a similar stellar mass and an over-massive DM halo \citep[$\sim 10^{12}\,\mathrm{M_\odot}$, see ][]{vanDokkum2016}.

A first possibility could be that DF2 had a larger stellar mass than today and gradually lost part of it due to stripping. This stripping would not modify the metallicity of DF2, but move its location in this plot horizontally towards lower stellar mass and thus closer to the relation. This process, which could explain the location of some dwarfs above the stellar mass-metallicity relation \citep[see the case of Antlia2, ][]{Torrealba2018}, could also move a metal-deficient UDG closer to the relation. Furthermore, \citet{Trujillo2018} note a significant brightening of DF2 in the Northern region in ultra-deep $g$-band Gemini data, which might be a trace of a past stripping event. We note that the stripping of the stars only begins when most of the DM mass is already lost \citep[around 90\%, see e.g.][]{Penarrubia2008}. Stellar mass stripping could then fit in the hypothesis of a DM-deficient galaxy. However, such a stripping scenario should also affect the GC system which should be stripped, or at least heated kinematically \citep{Smith2013}, which does not seem consistent with both the number (see next subsection) or the low velocity dispersion of the GCs associated with DF2.

A second possibility is that the gas of DF2 was already enriched in metals. This could be the case if DF2 was formed through tidally stripped material. We discuss this possibility in details in Sect.~\ref{sub:origin}.

Overall, DF2 shows a stellar population typical of quiescent UDGs. Its location in the mass-metallicity plane, which is very similar to that of dSphs, is not what one would expect for a DM-free galaxy. This could be a hint to the origin of this galaxy.

    \subsubsection{In terms of GC systems}
    
In Fig.~\ref{fig:age_Z} we saw that the metallicity of the GCs surrounding DF2 is significantly lower than that of the UDG, by around 0.5~dex. \citet{Lotz2004} found that field stars in 45 local dE are typically 0.1-0.2~mag redder than their GCs which they interpreted as a legacy of different star formation events and/or different metallicities. This color mismatch seems to be lower for UDGs \citep[less than 0.05~mag for DF17 and DF2][]{Beasley2016, vanDokkum2018_gc}. In the case of DF2, we can show that this is driven by the stellar body being more metal-rich than the GCs. This is typical for dwarf galaxies of similar masses, including the Fornax dSph which has an excess of GCs \citep[see e.g.][]{Cole2012, Larsen2014}. 

In the left panel of Fig.~\ref{fig:lick_comparison} we see that the $\alpha$-enrichment of the GCs is between [$\alpha$/Fe] = 0 and 0.3. These are also typical values for GCs in dwarfs, whose GCs are known to be less $\alpha$-enriched than those in more massive galaxies \citep{Sharina2010}. Thus the stellar populations of GCs around DF2 do not seem to deviate from previous known systems.

DF2 seems to have a rather high specific frequency\footnote{Number of GC (N$_\mathrm{GC}$) per 15 absolute magnitude in the V band ($M_V$): $S_N$ = N$_\mathrm{GC}~10^{0.4(M_V + 15)}$ \citep{Harris1981}.} of GCs compared to other UDGs \citep[above 11, see ][]{vanDokkum2018_gc}. Studies have shown that the $S_N$ of UDGs varies dramatically from galaxy to galaxy and is on average higher than in dwarf galaxies \citep{amorisco18, lim18}.

Moreover, we note that the fraction of light that is in GCs for DF2 is similar to that of other UDGs \citep[such as DF~17, see][]{vanDokkum2015, Peng2016}. The only feature by which the GC system of DF2 differs from other GC systems, and which remains unexplained, is that the peak magnitude of the GC luminosity function is unusually high if one assumes a 20~Mpc distance \citep{vanDokkum2018_gc}. We note that \citet{Trujillo2018} found that the GC luminosity function of DF2 is standard, if located at a distance of 13~Mpc.

\subsubsection{In terms of PNe}

It is the first time that PNe are discovered around a UDG, thanks to the use of an integral field unit (IFU) spectrograph with good spatial resolution. One may compare our number of detections with an estimate of the expected number of PNe for such a system.

The total number of PNe per bolometric luminosity of the host galaxy is parametrized as $\alpha = N_{PN} / L_{bol}$. We define $\alpha_{2.5}$ as the number of PNe in the brightest 2.5 mag of the PNLF per bolometric luminosity. While stellar evolutionary models still have difficulties in reproducing the constancy of the PNLF bright cut-off in galaxies of different morphology \citep[see e.g.\ ][]{Marigo2004}, the study of the luminosity-specific PN numbers (the $\alpha$ parameter) in external galaxies \citep{Buzzoni2006} provides a way of estimating the expected number of PNe in a galaxy. A typical $\alpha$ for metal-poor populations is $\sim 3 \times 10^{-7}$ PN per $L_{bol}/L_\odot$. The three detected PNe are probably in the brightest 2.5 mag of the PNLF, and, using the standard PNLF, $\alpha_{2.5} \approx \alpha/10$. So if $L_{bol} \approx 6\times 10^6$ to $10^8$ $L_\odot$ for DF2, then our 3 PNe imply $\alpha \sim 3 \times 10^{-8}$ to $5 \times 10^{-7}$, in reasonable agreement with expectations for a metal-poor stellar population \citep{Buzzoni2006}. Note that our field of view does not cover all the outskirts of DF2, where other PNe may be found. Thus DF2 does not seem to have a different PNe formation rate than other systems.

\subsection{What is the distance to DF2?}
\label{disc_dist}

The distance of DF2 is subject of a yet unsettled debate. Indeed, as noted in \citet{vanDokkum2018_nat}, a shorter distance would give a smaller stellar mass and increase the DM mass needed to recover the velocity dispersion measured. 

\citet{vanDokkum2018_nat} computed a distance of  $19.0 \pm 1.7$~Mpc from the surface brightness fluctuations (SBF) of the stellar body of DF2 and adopted a nominal distance of 20~Mpc. This distance was confirmed by an independent team, using the same technique \citep{Blakeslee2018}. \citet{Trujillo2018} claim that the calibration used by \citet{vanDokkum2018_nat} is only valid for colors redder than that of DF2, and that the extrapolation to bluer colors is not trivial. They use five different redshift-independent methods to compute the distance of DF2 which all give consistent result of $\sim$13~Mpc. For such a distance, the measured velocity dispersion cannot be achieved without a significant DM content. \citet{vanDokkum2018_dist} demonstrated that the tip of the red giant branch (TRGB) stars may be blended in the \textsc{HST} images. By using a megamaser-TRGB-SBF distance ladder they find a new estimate of the distance of 18.7$\pm$1.7~Mpc, which is consistent with their first distance estimate.

Another reliable distance estimator at these distances is the bright abrupt cut-off of the PNLF, whose absolute magnitude is almost independent of galaxy type, at around $M^\star = -4.51$~mag \citep[see ][for a recent review]{Ciardullo2012}. However, a trend towards a fainter cut-off magnitude in low metallicity galaxies is expected from theoretical models \citep{Dopita1992, Schonberner2010}, which is confirmed by observations \citep[see e.g.][]{Ciardullo2012}. Unfortunately, low-metallicity objects are usually not very massive and do not have enough PNe to sample well the PNLF. Hence, the metallicity dependence of $M^\star$ is hard to probe at the low metallicity end. In particular, the \citet{Dopita1992} theoretical relation was not confirmed at metallicities lower than that of the SMC. The cut-off magnitude of the low metallicity SMC and NGC~55 are estimated to be around M$^\star=-4.10$ \citep[see review by ][]{Ciardullo2012}. The SNR does not allow us to detect the [NII] line in the spectra of the PNe for a direct metallicity estimate of the PNe. If one extrapolates the  \citet{Dopita1992} relation to the stellar metallicity of 1/10th solar for DF2, derived in Section~\ref{sub:params}, one would expect a cut-off magnitude of M$^\star=-3.67$ for the PNLF of DF2.

   \begin{figure}
   \centering
   \includegraphics[angle=0,width=9cm]{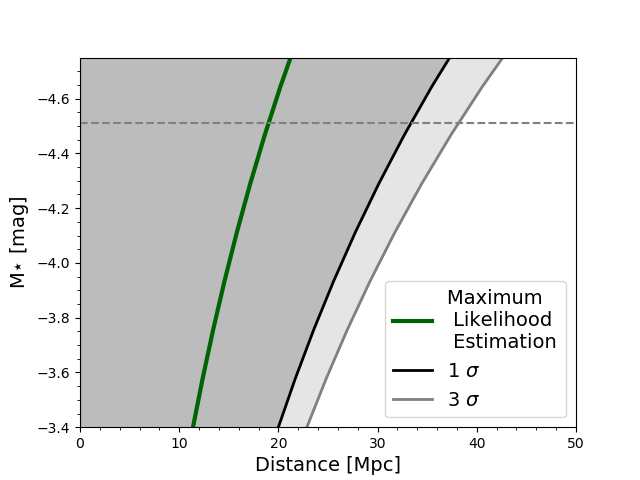}
      \caption{Maximum likelihood estimation and error range on the distance and cut-off magnitude. The distance which maximizes the likelihood is shown in green. The 1-$\sigma$ error sector is colored in dark gray with a thick dark line border. The 3-$\sigma$ error sector is colored in light gray with a thick gray line border. The horizontal dashed gray line shows M$^\star=-4.51$.
              }
         \label{fig:MLE}
   \end{figure}

Our IFU observations allowed us to find three PNe, which is a premi\`ere for UDGs. To quantify how much these three PNe inform us on the distance estimate, we perform a maximum likelihood estimation (MLE), using the PNLF from \citep{Ciardullo1989}. For a cut-off magnitude $M^\star$, the number of PN with absolute magnitude $M$ is proportional to:

\begin{equation}
    \mathrm{N}(M) \propto e^{0.307 M} \big(1 - e^{3(M^\star - M) } \big)
\end{equation}

The likelihood function L can be written:

\begin{equation}
    \mathrm{L} = \prod_{i = 1}^3 \frac{ \mathrm{N}(m_{i} - \mu) }  {\int_{M^\star}^{m_l- \mu} \mathrm{N}(m)~dm  } 
\end{equation}

with $\mu$ the distance modulus, $m_i$ the apparent magnitude of each PN, and $m_l$ the completeness limit. We set as completeness limit a [O~\textsc{iii}] emission line peaking at three times the local rms measured for the PNe. This gives $m_l =30.64$~mag. We minimize -ln(L) by varying $\mu$ and $M^\star$. We define respectively the 1-$\sigma$ and 3-$\sigma$ error range by the range of parameters for which respectively, $\Delta$ ln(L) < 0.5 and 4.5. In Fig.~\ref{fig:MLE} we show the result of the MLE.
For $M^\star=-4.51$, the distance that maximizes the likelihood is 19.0~Mpc, and the 1 and 3-$\sigma$ upper limits are respectively 33.3 and 38.1~Mpc. These values decrease for fainter M$^\star$. In particular, for M$\star=-3.67$, which is the value expected from \citet{Dopita1992} for 1/10th solar metallicity, the distance that maximizes the likelihood is 12.9~Mpc, and the 1 and 3-$\sigma$ upper limits are respectively 22.6 and 25.9~Mpc. \\

Thus, we note that none of the two former distance estimates is significantly more likely, given the three discovered PNe.
Given the off-centered field of view that we chose, we may have missed a brighter PN in the South-West part of DF2. In order to give strong constraints on the distance to DF2 the potential brightest PNe would need to be 1.5~mag brighter than PN1, magnitude for which a 20~Mpc distance would be ruled out by 3-$\sigma$.

\subsection{What is the origin of DF2?}
\label{sub:origin}

 From a kinematic study of 10 GCs surrounding DF2, \citet{vanDokkum2018_nat} inferred a low (projected) velocity dispersion, which they interpreted as DF2 `lacking' DM. This claim has been heavily scrutinized \citep{Trujillo2018, Famaey2018, Laporte2018,kroupa2018} and is revisited in Paper I. If confirmed by other independent tracers, this lack of DM calls for an additional formation channel, to explain the existence of both DM-deficient and DM-dominated UDGs \citep[such as DF44,][]{vanDokkum2016}. \\

As a first hypothesis, \citet{vanDokkum2018_nat} propose that the claimed `lack' of dark matter in DF2 may be explained if DF2 is a tidal dwarf galaxy (TDG), i.e. a galaxy formed from material that was expelled from a massive galaxy host during a galactic interaction \citep[see review by][]{Duc1999}. The proximity of the massive galaxy NGC~1052 and the peculiar radial velocity of DF2 (+293~km.s$^{-1}$ if at 20~Mpc) would support this hypothesis. Moreover, galaxies with typical morphological parameters of UDGs were observed to be still connected by a stellar stream to a massive host \citep{Bennet2018}. Unfortunately, no measurement of the stellar populations of those systems has been performed so far.

Because of this particular mode of formation, TDGs are indeed expected to be dark-matter free \citep{Bournaud2006, Wetzstein2007, Lelli15} and almost devoid of stars from their host \citep{Boquien2010}. Furthermore, old TDGs enter in the category of UDGs with their low central surface brightness and large effective radii \citep{Duc2014}. Interestingly, we note that the cluster formation efficiency of TDGs, that is the fraction of SFR which happens in bound clusters, is seen to be very high (50\%) compared to other systems (Fensch et al., subm.). Last but not least, they inherit the metal-enrichment from their more massive host. All observed tidal dwarf galaxies, which have stellar ages of typically less than 1~Gyr, deviate from the luminosity-metallicity diagram and have a significantly higher metallicity than other dwarfs for a similar luminosity, with a metallicity of typically around half solar, independent of their mass \citep{Duc2000, Weilbacher2003}. They are thus outliers of the stellar mass-metallicity relation.

Given the age of the stellar population of DF2, if it is a TDG the interaction must have happened at around $z$ = 2, where the metal-enrichment of the gas in the outskirts of the host galaxy could still be quite low \cite[see e.g.][]{Jones2013}. Unfortunately, there is not much data on old TDGs as their low surface brightness makes them difficult to study \citep[but see][]{Duc2014}, unless some or all of the Milky Way and Andromeda satellite galaxies are very old TDGs \citep{Metz2007, Pawlowski2011,Yang2014}. In Sect.~\ref{sub:disc_stel_pop}, we noted that DF2 is on the stellar mass-metallicity relation, contrarily to the young TDGs. If the level of pre-enrichment is between 0.001 and 0.01 Z$_\odot$, it is possible that the TDGs would not reach the mass-metallicity relation of young TDGs after many Gyrs \citep{Recchi2015}. Under the DM-deficiency hypothesis, a small pre-enrichment of DF2 could then explain the location of DF2 in the stellar mass-metallicity diagram (see discussion in Sect.~\ref{sub:disc_stel_pop}). Moreover, we note that most GCs in 'normal' dwarf galaxies with spectroscopic metallicity measurements are very metal-poor (e.g. [Fe/H]$\sim -$2 dex, for  GCs in the Fornax dSph, see \citealt{deBoer2016}). A coeval formation of the UDG and its clusters in pre-enriched gas ejected from a massive galaxy could explain how the GCs of DF2 have been enriched to [Fe/H]$\sim -$1.6~dex. Thus the metallicity of DF2 and its GCs could be consistent with the TDG origin hypothesis. 

\section{Conclusions}
\label{conclusion}

We present the first simultaneous analysis of the stellar population of a UDG and its surrounding globular clusters. 

We fit SSPs to the starlight component of the stellar body and the stack of all GCs using the empirical stellar library eMILES with the fitting routine pPXF.

We find that the UDG's stellar populations are consistent with an old age, 8.9$\pm$1.5~Gyr, low metallicity, [M/H]~$ = -1.19 \pm 0.11$, and little to no $\alpha$-enrichment, i.e formed over a timescale larger than 1~Gyr. The GC spectra are consistent with the same age, 8.9 $\pm$ 1.4~Gyr, but have a lower metallicity than DF2 ([Fe/H]~$= -1.55 \pm 0.09$). This result is consistent with the Lick indices diagnostics and the broadband colors of DF2 and its clusters.

The stellar mass and metallicity of the UDG fall on the empirical relation found for old dwarf galaxies. In particular, DF2 has a comparable metallicity to DF44, which has the same stellar mass but was shown to have a MW-like DM-halo. This relation is a consequence of the self-enrichment of galaxies and thus depend on the total mass of the galaxy. Under the DM-deficiency hypothesis one would then expect DF2 to have lower metallicity than galaxies with similar stellar mass. We note that stellar mass loss due to stripping could move a metal-deficient galaxy back to the relation, but this would affect its GC system, which does not seem to be the case for DF2. Another hypothesis would be that DF2 has a tidal origin and was formed by gas pre-enriched in metals.

We also report the discovery of the first three PNe in a UDG. That number is consistent with the number of PNe in other galaxies with similar luminosity and metallicities. We find that distance estimates of 13 to 20 Mpc are similarly likely, given the three discovered PNe.

\begin{acknowledgements}
We thank the anonymous referee for their constructive comments which helped improve this paper. We thank Lodovico Coccato for his extremely useful help on the use of \textsc{ZAP}. We thank Simon Conseil for his very reactive answers to our questions on \textsc{ZAP}. We thank Harald Kuntschner for useful discussions on the measurement and interpretation of Lick indices, and for letting us use his code. We thank Magda Arnaboldi for very useful discussions on PNe. We thank Souradeep Bhattacharya and Glenn van de Ven
for helpful discussions. OM is grateful to the Swiss National Science Foundation for financial support. E.W.P. acknowledges support from the National Natural Science Foundation of China through Grant No. 11573002 \\
\end{acknowledgements}

\bibliographystyle{aa}  
\bibliography{library}

\begin{thebibliography}{95}
\expandafter\ifx\csname natexlab\endcsname\relax\def\natexlab#1{#1}\fi

\bibitem[{{Allen}(1973)}]{Allen1973}
{Allen}, C.~W. 1973, {Astrophysical quantities}

\bibitem[{{Amorisco} \& {Loeb}(2016)}]{Amorisco2016}
{Amorisco}, N.~C. \& {Loeb}, A. 2016, \mnras, 459, L51

\bibitem[{{Amorisco} {et~al.}(2018){Amorisco}, {Monachesi}, {Agnello}, \&
  {White}}]{amorisco18}
{Amorisco}, N.~C., {Monachesi}, A., {Agnello}, A., \& {White}, S.~D.~M. 2018,
  \mnras, 475, 4235

\bibitem[{{Beasley} {et~al.}(2016){Beasley}, {Romanowsky}, {Pota}, {Navarro},
  {Martinez Delgado}, {Neyer}, \& {Deich}}]{Beasley2016}
{Beasley}, M.~A., {Romanowsky}, A.~J., {Pota}, V., {et~al.} 2016, \apjl, 819,
  L20

\bibitem[{{Beasley} \& {Trujillo}(2016)}]{beasley2016tr}
{Beasley}, M.~A. \& {Trujillo}, I. 2016, \apj, 830, 23

\bibitem[{{Bennet} {et~al.}(2018){Bennet}, {Sand}, {Zaritsky}, {Crnojevi{\'c}},
  {Spekkens}, \& {Karunakaran}}]{Bennet2018}
{Bennet}, P., {Sand}, D.~J., {Zaritsky}, D., {et~al.} 2018, \apjl, 866, L11

\bibitem[{{Blakeslee} \& {Cantiello}(2018)}]{Blakeslee2018}
{Blakeslee}, J.~P. \& {Cantiello}, M. 2018, Research Notes of the American
  Astronomical Society, 2, 146

\bibitem[{{Blakeslee} {et~al.}(1997){Blakeslee}, {Tonry}, \&
  {Metzger}}]{Blakeslee1997}
{Blakeslee}, J.~P., {Tonry}, J.~L., \& {Metzger}, M.~R. 1997, \aj, 114, 482

\bibitem[{{Boquien} {et~al.}(2010){Boquien}, {Duc}, {Galliano}, {Braine},
  {Lisenfeld}, {Charmandaris}, \& {Appleton}}]{Boquien2010}
{Boquien}, M., {Duc}, P.-A., {Galliano}, F., {et~al.} 2010, \aj, 140, 2124

\bibitem[{{Bournaud} \& {Duc}(2006)}]{Bournaud2006}
{Bournaud}, F. \& {Duc}, P.-A. 2006, \aap, 456, 481

\bibitem[{{Buzzoni} {et~al.}(2006){Buzzoni}, {Arnaboldi}, \&
  {Corradi}}]{Buzzoni2006}
{Buzzoni}, A., {Arnaboldi}, M., \& {Corradi}, R.~L.~M. 2006, \mnras, 368, 877

\bibitem[{{Cappellari}(2017)}]{Cappellari2017}
{Cappellari}, M. 2017, \mnras, 466, 798

\bibitem[{{Cappellari} \& {Emsellem}(2004)}]{Cappellari2004}
{Cappellari}, M. \& {Emsellem}, E. 2004, \pasp, 116, 138

\bibitem[{{Chan} {et~al.}(2018){Chan}, {Kere{\v s}}, {Wetzel}, {Hopkins},
  {Faucher-Gigu{\`e}re}, {El-Badry}, {Garrison-Kimmel}, \&
  {Boylan-Kolchin}}]{Chan2018}
{Chan}, T.~K., {Kere{\v s}}, D., {Wetzel}, A., {et~al.} 2018, \mnras, 478, 906

\bibitem[{{Chowdhury}(2019)}]{Chowdhury2018}
{Chowdhury}, A. 2019, \mnras, 482, L99

\bibitem[{{Ciardullo}(2012)}]{Ciardullo2012}
{Ciardullo}, R. 2012, \apss, 341, 151

\bibitem[{{Ciardullo} {et~al.}(1989){Ciardullo}, {Jacoby}, {Ford}, \&
  {Neill}}]{Ciardullo1989}
{Ciardullo}, R., {Jacoby}, G.~H., {Ford}, H.~C., \& {Neill}, J.~D. 1989, \apj,
  339, 53

\bibitem[{{Cole} {et~al.}(2012){Cole}, {Dehnen}, {Read}, \&
  {Wilkinson}}]{Cole2012}
{Cole}, D.~R., {Dehnen}, W., {Read}, J.~I., \& {Wilkinson}, M.~I. 2012, \mnras,
  426, 601

\bibitem[{{Collins} {et~al.}(2013){Collins}, {Chapman}, {Rich}, {Ibata},
  {Martin}, {Irwin}, {Bate}, {Lewis}, {Pe{\~n}arrubia}, {Arimoto}, {Casey},
  {Ferguson}, {Koch}, {McConnachie}, \& {Tanvir}}]{Collins2013}
{Collins}, M.~L.~M., {Chapman}, S.~C., {Rich}, R.~M., {et~al.} 2013, \apj, 768,
  172

\bibitem[{{Conroy} {et~al.}(2009){Conroy}, {Gunn}, \& {White}}]{Conroy2009}
{Conroy}, C., {Gunn}, J.~E., \& {White}, M. 2009, \apj, 699, 486

\bibitem[{{Conroy} {et~al.}(2018){Conroy}, {Villaume}, {van Dokkum}, \&
  {Lind}}]{Conroy2018}
{Conroy}, C., {Villaume}, A., {van Dokkum}, P.~G., \& {Lind}, K. 2018, \apj,
  854, 139

\bibitem[{{Conselice} {et~al.}(2003){Conselice}, {Gallagher}, \&
  {Wyse}}]{conselice03}
{Conselice}, C.~J., {Gallagher}, III, J.~S., \& {Wyse}, R.~F.~G. 2003, \aj,
  125, 66

\bibitem[{{Dalcanton} {et~al.}(1997){Dalcanton}, {Spergel}, {Gunn}, {Schmidt},
  \& {Schneider}}]{dalcanton97}
{Dalcanton}, J.~J., {Spergel}, D.~N., {Gunn}, J.~E., {Schmidt}, M., \&
  {Schneider}, D.~P. 1997, \aj, 114, 635

\bibitem[{{de Boer} \& {Fraser}(2016)}]{deBoer2016}
{de Boer}, T.~J.~L. \& {Fraser}, M. 2016, \aap, 590, A35

\bibitem[{{Di Cintio} {et~al.}(2017){Di Cintio}, {Brook}, {Dutton},
  {Macci{\`o}}, {Obreja}, \& {Dekel}}]{dicintio17}
{Di Cintio}, A., {Brook}, C.~B., {Dutton}, A.~A., {et~al.} 2017, \mnras, 466,
  L1

\bibitem[{{Dopita} {et~al.}(1992){Dopita}, {Jacoby}, \&
  {Vassiliadis}}]{Dopita1992}
{Dopita}, M.~A., {Jacoby}, G.~H., \& {Vassiliadis}, E. 1992, \apj, 389, 27

\bibitem[{{Duc} {et~al.}(2000){Duc}, {Brinks}, {Springel}, {Pichardo},
  {Weilbacher}, \& {Mirabel}}]{Duc2000}
{Duc}, P.-A., {Brinks}, E., {Springel}, V., {et~al.} 2000, \aj, 120, 1238

\bibitem[{{Duc} \& {Mirabel}(1999)}]{Duc1999}
{Duc}, P.-A. \& {Mirabel}, I.~F. 1999, in IAU Symposium, Vol. 186, Galaxy
  Interactions at Low and High Redshift, ed. J.~E. {Barnes} \& D.~B. {Sanders},
  61

\bibitem[{{Duc} {et~al.}(2014){Duc}, {Paudel}, {McDermid}, {Cuillandre},
  {Serra}, {Bournaud}, {Cappellari}, \& {Emsellem}}]{Duc2014}
{Duc}, P.-A., {Paudel}, S., {McDermid}, R.~M., {et~al.} 2014, \mnras, 440, 1458

\bibitem[{{Evstigneeva} {et~al.}(2007){Evstigneeva}, {Gregg}, {Drinkwater}, \&
  {Hilker}}]{evstigneeva07}
{Evstigneeva}, E.~A., {Gregg}, M.~D., {Drinkwater}, M.~J., \& {Hilker}, M.
  2007, \aj, 133, 1722

\bibitem[{{Famaey} {et~al.}(2018){Famaey}, {McGaugh}, \&
  {Milgrom}}]{Famaey2018}
{Famaey}, B., {McGaugh}, S., \& {Milgrom}, M. 2018, \mnras, 480, 473

\bibitem[{{Ferr{\'e}-Mateu} {et~al.}(2018){Ferr{\'e}-Mateu}, {Alabi}, {Forbes},
  {Romanowsky}, {Brodie}, {Pandya}, {Mart{\'{\i}}n-Navarro}, {Bellstedt},
  {Wasserman}, {Stone}, \& {Okabe}}]{Ferre-Mateu2018}
{Ferr{\'e}-Mateu}, A., {Alabi}, A., {Forbes}, D.~A., {et~al.} 2018, \mnras,
  479, 4891

\bibitem[{{Gallazzi} {et~al.}(2005){Gallazzi}, {Charlot}, {Brinchmann},
  {White}, \& {Tremonti}}]{Gallazzi2005}
{Gallazzi}, A., {Charlot}, S., {Brinchmann}, J., {White}, S.~D.~M., \&
  {Tremonti}, C.~A. 2005, \mnras, 362, 41

\bibitem[{{Girardi} {et~al.}(2000){Girardi}, {Bressan}, {Bertelli}, \&
  {Chiosi}}]{Girardi2000}
{Girardi}, L., {Bressan}, A., {Bertelli}, G., \& {Chiosi}, C. 2000, \aaps, 141,
  371

\bibitem[{{Gu} {et~al.}(2018){Gu}, {Conroy}, {Law}, {van Dokkum}, {Yan},
  {Wake}, {Bundy}, {Merritt}, {Abraham}, {Zhang}, {Bershady}, {Bizyaev},
  {Brinkmann}, {Drory}, {Grabowski}, {Masters}, {Pan}, {Parejko}, {Weijmans},
  \& {Zhang}}]{Gu2018}
{Gu}, M., {Conroy}, C., {Law}, D., {et~al.} 2018, \apj, 859, 37

\bibitem[{{Harris} {et~al.}(2017){Harris}, {Blakeslee}, \& {Harris}}]{harris17}
{Harris}, W.~E., {Blakeslee}, J.~P., \& {Harris}, G.~L.~H. 2017, \apj, 836, 67

\bibitem[{{Harris} \& {van den Bergh}(1981)}]{Harris1981}
{Harris}, W.~E. \& {van den Bergh}, S. 1981, \aj, 86, 1627

\bibitem[{{Impey} {et~al.}(1988){Impey}, {Bothun}, \& {Malin}}]{impey88}
{Impey}, C., {Bothun}, G., \& {Malin}, D. 1988, \apj, 330, 634

\bibitem[{{Jones} {et~al.}(2013){Jones}, {Ellis}, {Richard}, \&
  {Jullo}}]{Jones2013}
{Jones}, T., {Ellis}, R.~S., {Richard}, J., \& {Jullo}, E. 2013, \apj, 765, 48

\bibitem[{{Kadowaki} {et~al.}(2017){Kadowaki}, {Zaritsky}, \&
  {Donnerstein}}]{Kadowaki2017}
{Kadowaki}, J., {Zaritsky}, D., \& {Donnerstein}, R.~L. 2017, \apjl, 838, L21

\bibitem[{{Karachentsev} {et~al.}(2000){Karachentsev}, {Karachentseva},
  {Suchkov}, \& {Grebel}}]{Karachentsev2000}
{Karachentsev}, I.~D., {Karachentseva}, V.~E., {Suchkov}, A.~A., \& {Grebel},
  E.~K. 2000, \aaps, 145, 415

\bibitem[{{Kirby} {et~al.}(2013){Kirby}, {Cohen}, {Guhathakurta}, {Cheng},
  {Bullock}, \& {Gallazzi}}]{Kirby2013}
{Kirby}, E.~N., {Cohen}, J.~G., {Guhathakurta}, P., {et~al.} 2013, \apj, 779,
  102

\bibitem[{{Koda} {et~al.}(2015){Koda}, {Yagi}, {Yamanoi}, \&
  {Komiyama}}]{koda15}
{Koda}, J., {Yagi}, M., {Yamanoi}, H., \& {Komiyama}, Y. 2015, \apjl, 807, L2

\bibitem[{{K{\"o}ppen} {et~al.}(2007){K{\"o}ppen}, {Weidner}, \&
  {Kroupa}}]{Koeppen2007}
{K{\"o}ppen}, J., {Weidner}, C., \& {Kroupa}, P. 2007, \mnras, 375, 673

\bibitem[{{Kroupa}(2001)}]{Kroupa2001}
{Kroupa}, P. 2001, \mnras, 322, 231

\bibitem[{{Kroupa}(2012)}]{Kroupa2012}
{Kroupa}, P. 2012, \pasa, 29, 395

\bibitem[{{Kroupa} {et~al.}(2018){Kroupa}, {Haghi}, {Javanmardi}, {Zonoozi},
  {M{\"u}ller}, {Banik}, {Wu}, {Zhao}, \& {Dabringhausen}}]{kroupa2018}
{Kroupa}, P., {Haghi}, H., {Javanmardi}, B., {et~al.} 2018, \nat, 561, E4

\bibitem[{{Kuntschner} {et~al.}(2010){Kuntschner}, {Emsellem}, {Bacon},
  {Cappellari}, {Davies}, {de Zeeuw}, {Falc{\'o}n-Barroso}, {Krajnovi{\'c}},
  {McDermid}, {Peletier}, {Sarzi}, {Shapiro}, {van den Bosch}, \& {van de
  Ven}}]{Kuntschner2010}
{Kuntschner}, H., {Emsellem}, E., {Bacon}, R., {et~al.} 2010, \mnras, 408, 97

\bibitem[{{Laporte} {et~al.}(2018){Laporte}, {Agnello}, \&
  {Navarro}}]{Laporte2018}
{Laporte}, C.~F.~P., {Agnello}, A., \& {Navarro}, J.~F. 2018, MNRAS in press,
  [\eprint[arXiv]{1804.04139}]

\bibitem[{{Larsen} {et~al.}(2014){Larsen}, {Brodie}, {Forbes}, \&
  {Strader}}]{Larsen2014}
{Larsen}, S.~S., {Brodie}, J.~P., {Forbes}, D.~A., \& {Strader}, J. 2014, \aap,
  565, A98

\bibitem[{{Lelli} {et~al.}(2015){Lelli}, {Duc}, {Brinks}, {Bournaud},
  {McGaugh}, {Lisenfeld}, {Weilbacher}, {Boquien}, {Revaz}, {Braine},
  {Koribalski}, \& {Belles}}]{Lelli15}
{Lelli}, F., {Duc}, P.-A., {Brinks}, E., {et~al.} 2015, \aap, 584, A113

\bibitem[{{Lim} {et~al.}(2018){Lim}, {Peng}, {C{\^o}t{\'e}}, {Sales}, {den
  Brok}, {Blakeslee}, \& {Guhathakurta}}]{lim18}
{Lim}, S., {Peng}, E.~W., {C{\^o}t{\'e}}, P., {et~al.} 2018, \apj, 862, 82

\bibitem[{{Lotz} {et~al.}(2004){Lotz}, {Miller}, \& {Ferguson}}]{Lotz2004}
{Lotz}, J.~M., {Miller}, B.~W., \& {Ferguson}, H.~C. 2004, \apj, 613, 262

\bibitem[{{Marigo} {et~al.}(2004){Marigo}, {Girardi}, {Weiss}, {Groenewegen},
  \& {Chiosi}}]{Marigo2004}
{Marigo}, P., {Girardi}, L., {Weiss}, A., {Groenewegen}, M.~A.~T., \& {Chiosi},
  C. 2004, \aap, 423, 995

\bibitem[{{Martin} {et~al.}(2018){Martin}, {Collins}, {Longeard}, \&
  {Tollerud}}]{Martin2018}
{Martin}, N.~F., {Collins}, M.~L.~M., {Longeard}, N., \& {Tollerud}, E. 2018,
  \apjl, 859, L5

\bibitem[{{Metz} \& {Kroupa}(2007)}]{Metz2007}
{Metz}, M. \& {Kroupa}, P. 2007, \mnras, 376, 387

\bibitem[{{Mihos} {et~al.}(2015){Mihos}, {Durrell}, {Ferrarese}, {Feldmeier},
  {C{\^o}t{\'e}}, {Peng}, {Harding}, {Liu}, {Gwyn}, \& {Cuillandre}}]{mihos15}
{Mihos}, J.~C., {Durrell}, P.~R., {Ferrarese}, L., {et~al.} 2015, \apjl, 809,
  L21

\bibitem[{{Mu{\~n}oz} {et~al.}(2015){Mu{\~n}oz}, {Eigenthaler}, {Puzia},
  {Taylor}, {Ordenes-Brice{\~n}o}, {Alamo-Mart{\'{\i}}nez}, {Ribbeck},
  {{\'A}ngel}, {Capaccioli}, {C{\^o}t{\'e}}, {Ferrarese}, {Galaz}, {Hempel},
  {Hilker}, {Jord{\'a}n}, {Lan{\c c}on}, {Mieske}, {Paolillo}, {Richtler},
  {S{\'a}nchez-Janssen}, \& {Zhang}}]{munoz15}
{Mu{\~n}oz}, R.~P., {Eigenthaler}, P., {Puzia}, T.~H., {et~al.} 2015, \apjl,
  813, L15

\bibitem[{{M{\"u}ller} {et~al.}(2018){M{\"u}ller}, {Jerjen}, \&
  {Binggeli}}]{Mueller18}
{M{\"u}ller}, O., {Jerjen}, H., \& {Binggeli}, B. 2018, \aap, 615, A105

\bibitem[{{Pandya} {et~al.}(2018){Pandya}, {Romanowsky}, {Laine}, {Brodie},
  {Johnson}, {Glaccum}, {Villaume}, {Cuillandre}, {Gwyn}, {Krick}, {Lasker},
  {Mart{\'{\i}}n-Navarro}, {Martinez-Delgado}, \& {van Dokkum}}]{Pandya2018}
{Pandya}, V., {Romanowsky}, A.~J., {Laine}, S., {et~al.} 2018, \apj, 858, 29

\bibitem[{{Pawlowski} {et~al.}(2011){Pawlowski}, {Kroupa}, \& {de
  Boer}}]{Pawlowski2011}
{Pawlowski}, M.~S., {Kroupa}, P., \& {de Boer}, K.~S. 2011, \aap, 532, A118

\bibitem[{{Pe{\~n}arrubia} {et~al.}(2008){Pe{\~n}arrubia}, {McConnachie}, \&
  {Navarro}}]{Penarrubia2008}
{Pe{\~n}arrubia}, J., {McConnachie}, A.~W., \& {Navarro}, J.~F. 2008, \apj,
  672, 904

\bibitem[{{Peng} {et~al.}(2004){Peng}, {Ford}, \& {Freeman}}]{peng2004}
{Peng}, E.~W., {Ford}, H.~C., \& {Freeman}, K.~C. 2004, \apjs, 150, 367

\bibitem[{{Peng} \& {Lim}(2016)}]{Peng2016}
{Peng}, E.~W. \& {Lim}, S. 2016, \apjl, 822, L31

\bibitem[{{Read} {et~al.}(2017){Read}, {Iorio}, {Agertz}, \&
  {Fraternali}}]{Read2017}
{Read}, J.~I., {Iorio}, G., {Agertz}, O., \& {Fraternali}, F. 2017, \mnras,
  467, 2019

\bibitem[{{Recchi} {et~al.}(2015){Recchi}, {Kroupa}, \&
  {Ploeckinger}}]{Recchi2015}
{Recchi}, S., {Kroupa}, P., \& {Ploeckinger}, S. 2015, \mnras, 450, 2367

\bibitem[{{Rom{\'a}n} \& {Trujillo}(2017)}]{roman17}
{Rom{\'a}n}, J. \& {Trujillo}, I. 2017, \mnras, 468, 4039

\bibitem[{{Rong} {et~al.}(2017){Rong}, {Guo}, {Gao}, {Liao}, {Xie}, {Puzia},
  {Sun}, \& {Pan}}]{Rong2017}
{Rong}, Y., {Guo}, Q., {Gao}, L., {et~al.} 2017, \mnras, 470, 4231

\bibitem[{{Ruiz-Lara} {et~al.}(2018){Ruiz-Lara}, {Beasley},
  {Falc{\'o}n-Barroso}, {Rom{\'a}n}, {Pinna}, {Brook}, {Cintio},
  {Mart{\'{\i}}n-Navarro}, {Trujillo}, \& {Vazdekis}}]{Ruiz-Lara2018}
{Ruiz-Lara}, T., {Beasley}, M.~A., {Falc{\'o}n-Barroso}, J., {et~al.} 2018,
  \mnras, 478, 2034

\bibitem[{{Sandage} \& {Binggeli}(1984)}]{sandage1984}
{Sandage}, A. \& {Binggeli}, B. 1984, \aj, 89, 919

\bibitem[{{Schiavon}(2007)}]{Schiavon2007}
{Schiavon}, R.~P. 2007, \apjs, 171, 146

\bibitem[{{Schlafly} \& {Finkbeiner}(2011)}]{Schlafly2011}
{Schlafly}, E.~F. \& {Finkbeiner}, D.~P. 2011, \apj, 737, 103

\bibitem[{{Schlegel} {et~al.}(1998){Schlegel}, {Finkbeiner}, \&
  {Davis}}]{Schlegel1998}
{Schlegel}, D.~J., {Finkbeiner}, D.~P., \& {Davis}, M. 1998, \apj, 500, 525

\bibitem[{{Sch{\"o}nberner} {et~al.}(2010){Sch{\"o}nberner}, {Jacob}, {Sandin},
  \& {Steffen}}]{Schonberner2010}
{Sch{\"o}nberner}, D., {Jacob}, R., {Sandin}, C., \& {Steffen}, M. 2010, \aap,
  523, A86

\bibitem[{{Sharina} {et~al.}(2010){Sharina}, {Chandar}, {Puzia}, {Goudfrooij},
  \& {Davoust}}]{Sharina2010}
{Sharina}, M.~E., {Chandar}, R., {Puzia}, T.~H., {Goudfrooij}, P., \&
  {Davoust}, E. 2010, \mnras, 405, 839

\bibitem[{{Shi} {et~al.}(2017){Shi}, {Zheng}, {Zhao}, {Pan}, {Li}, {Zou},
  {Zhou}, {Guo}, {An}, \& {Li}}]{Shi2017}
{Shi}, D.~D., {Zheng}, X.~Z., {Zhao}, H.~B., {et~al.} 2017, \apj, 846, 26

\bibitem[{{Sif{\'o}n} {et~al.}(2018){Sif{\'o}n}, {van der Burg}, {Hoekstra},
  {Muzzin}, \& {Herbonnet}}]{Sifon2018}
{Sif{\'o}n}, C., {van der Burg}, R.~F.~J., {Hoekstra}, H., {Muzzin}, A., \&
  {Herbonnet}, R. 2018, \mnras, 473, 3747

\bibitem[{{Smith} {et~al.}(2013){Smith}, {S{\'a}nchez-Janssen}, {Fellhauer},
  {Puzia}, {Aguerri}, \& {Farias}}]{Smith2013}
{Smith}, R., {S{\'a}nchez-Janssen}, R., {Fellhauer}, M., {et~al.} 2013, \mnras,
  429, 1066

\bibitem[{{Soto} {et~al.}(2016){Soto}, {Lilly}, {Bacon}, {Richard}, \&
  {Conseil}}]{Soto2016}
{Soto}, K.~T., {Lilly}, S.~J., {Bacon}, R., {Richard}, J., \& {Conseil}, S.
  2016, \mnras, 458, 3210

\bibitem[{{Thomas} {et~al.}(2010){Thomas}, {Maraston}, {Schawinski}, {Sarzi},
  \& {Silk}}]{thomas10}
{Thomas}, D., {Maraston}, C., {Schawinski}, K., {Sarzi}, M., \& {Silk}, J.
  2010, \mnras, 404, 1775

\bibitem[{{Torrealba} {et~al.}(2018){Torrealba}, {Belokurov}, {Koposov}, {Li},
  {Walker}, {Sanders}, {Geringer-Sameth}, {Zucker}, {Kuehn}, {Evans}, \&
  {Dehnen}}]{Torrealba2018}
{Torrealba}, G., {Belokurov}, V., {Koposov}, S.~E., {et~al.} 2018, arXiv
  e-prints [\eprint[arXiv]{1811.04082}]

\bibitem[{{Trujillo} {et~al.}(2018){Trujillo}, {Beasley}, {Borlaff},
  {Carrasco}, {Di Cintio}, {Filho}, {Monelli}, {Montes}, {Roman}, {Ruiz-Lara},
  {Sanchez Almeida}, {Valls-Gabaud}, \& {Vazdekis}}]{Trujillo2018}
{Trujillo}, I., {Beasley}, M.~A., {Borlaff}, A., {et~al.} 2018, ArXiv e-prints
  [\eprint[arXiv]{1806.10141}]

\bibitem[{{van der Burg} {et~al.}(2017){van der Burg}, {Hoekstra}, {Muzzin},
  {Sif{\'o}n}, {Viola}, {Bremer}, {Brough}, {Driver}, {Erben}, {Heymans},
  {Hildebrandt}, {Holwerda}, {Klaes}, {Kuijken}, {McGee}, {Nakajima},
  {Napolitano}, {Norberg}, {Taylor}, \& {Valentijn}}]{vanderBurg17}
{van der Burg}, R.~F.~J., {Hoekstra}, H., {Muzzin}, A., {et~al.} 2017, \aap,
  607, A79

\bibitem[{{van der Burg} {et~al.}(2016){van der Burg}, {Muzzin}, \&
  {Hoekstra}}]{vanderBurg2016}
{van der Burg}, R.~F.~J., {Muzzin}, A., \& {Hoekstra}, H. 2016, \aap, 590, A20

\bibitem[{{van Dokkum} {et~al.}(2016){van Dokkum}, {Abraham}, {Brodie},
  {Conroy}, {Danieli}, {Merritt}, {Mowla}, {Romanowsky}, \&
  {Zhang}}]{vanDokkum2016}
{van Dokkum}, P., {Abraham}, R., {Brodie}, J., {et~al.} 2016, \apjl, 828, L6

\bibitem[{{van Dokkum} {et~al.}(2018{\natexlab{a}}){van Dokkum}, {Cohen},
  {Danieli}, {Kruijssen}, {Romanowsky}, {Merritt}, {Abraham}, {Brodie},
  {Conroy}, {Lokhorst}, {Mowla}, {O'Sullivan}, \& {Zhang}}]{vanDokkum2018_gc}
{van Dokkum}, P., {Cohen}, Y., {Danieli}, S., {et~al.} 2018{\natexlab{a}},
  \apjl, 856, L30

\bibitem[{{van Dokkum} {et~al.}(2018{\natexlab{b}}){van Dokkum}, {Danieli},
  {Cohen}, {Merritt}, {Romanowsky}, {Abraham}, {Brodie}, {Conroy}, {Lokhorst},
  {Mowla}, {O'Sullivan}, \& {Zhang}}]{vanDokkum2018_nat}
{van Dokkum}, P., {Danieli}, S., {Cohen}, Y., {et~al.} 2018{\natexlab{b}},
  \nat, 555, 629

\bibitem[{{van Dokkum} {et~al.}(2018{\natexlab{c}}){van Dokkum}, {Danieli},
  {Cohen}, {Romanowsky}, \& {Conroy}}]{vanDokkum2018_dist}
{van Dokkum}, P., {Danieli}, S., {Cohen}, Y., {Romanowsky}, A.~J., \& {Conroy},
  C. 2018{\natexlab{c}}, \apjl, 864, L18

\bibitem[{{van Dokkum} {et~al.}(2015){van Dokkum}, {Abraham}, {Merritt},
  {Zhang}, {Geha}, \& {Conroy}}]{vanDokkum2015}
{van Dokkum}, P.~G., {Abraham}, R., {Merritt}, A., {et~al.} 2015, \apjl, 798,
  L45

\bibitem[{{Vazdekis} {et~al.}(2016){Vazdekis}, {Koleva}, {Ricciardelli},
  {R{\"o}ck}, \& {Falc{\'o}n-Barroso}}]{Vazdekis2016}
{Vazdekis}, A., {Koleva}, M., {Ricciardelli}, E., {R{\"o}ck}, B., \&
  {Falc{\'o}n-Barroso}, J. 2016, \mnras, 463, 3409

\bibitem[{{Weilbacher} {et~al.}(2003){Weilbacher}, {Duc}, \&
  {Fritze-v.~Alvensleben}}]{Weilbacher2003}
{Weilbacher}, P.~M., {Duc}, P.-A., \& {Fritze-v.~Alvensleben}, U. 2003, \aap,
  397, 545

\bibitem[{{Wetzstein} {et~al.}(2007){Wetzstein}, {Naab}, \&
  {Burkert}}]{Wetzstein2007}
{Wetzstein}, M., {Naab}, T., \& {Burkert}, A. 2007, \mnras, 375, 805

\bibitem[{{Worthey} {et~al.}(1994){Worthey}, {Faber}, {Gonzalez}, \&
  {Burstein}}]{worthey94}
{Worthey}, G., {Faber}, S.~M., {Gonzalez}, J.~J., \& {Burstein}, D. 1994,
  \apjs, 94, 687

\bibitem[{{Yang} {et~al.}(2014){Yang}, {Hammer}, {Fouquet}, {Flores}, {Puech},
  {Pawlowski}, \& {Kroupa}}]{Yang2014}
{Yang}, Y., {Hammer}, F., {Fouquet}, S., {et~al.} 2014, \mnras, 442, 2419

\bibitem[{{Yozin} \& {Bekki}(2015)}]{Yozin2015}
{Yozin}, C. \& {Bekki}, K. 2015, \mnras, 452, 937

\end{thebibliography}

\end{document}